\documentclass[twocolumn,prb,showpacs,superscriptaddress]{revtex4-1}
\usepackage{amsfonts}
\usepackage{graphicx}
\usepackage{amsmath}
\usepackage{times}
\usepackage{amssymb}
\usepackage{color}
\usepackage{IEEEtrantools}
\usepackage{subfigure}
\usepackage{verbatim}
\usepackage{siunitx}
\usepackage{braket}
\usepackage[version=3]{mhchem}
\usepackage[colorlinks,bookmarks=false,citecolor=blue,linkcolor=red,urlcolor=blue]{hyperref}

\newenvironment{eqns}[1][rCl]
  {\begin{IEEEeqnarray}{#1}}
  {\end{IEEEeqnarray}\ignorespacesafterend}

\renewcommand {\i} { \mathrm{i} }               
\newcommand {\e}{ \mathrm{e} }               
\newcommand {\dx} {\; \mathrm{d} }   

\begin{document}

\title{Magnetic chern bands and triplon Hall effect in an extended Shastry-Sutherland model}

\author{M. Malki}
\affiliation{Lehrstuhl f\"ur Theoretische Physik I, Otto-Hahn-Str.~4, TU Dortmund, D-44221 Dortmund, Germany}
\email{maik.malki@tu-dortmund.de}
\author{K.P. Schmidt}
\affiliation{Institut f\"ur Theoretische Physik, Universit\"at Erlangen-N\"urnberg, Staudtstra\ss e 7, 91058 Erlangen, Germany}
\email{kai.phillip.schmidt@fau.de}

\date{\today}

\begin{abstract}
We study topological properties of one-triplon bands in an extended Shastry-Sutherland model relevant for the frustrated quantum magnet SrCu$_2$(BO$_3$)$_2$. To this end perturbative continuous unitary transformations are applied about the isolated dimer limit allowing to calculate the one-triplon dispersion up to high order in various couplings including intra and inter Dzyaloshinskii-Moriya interactions and a general uniform magnetic field. We determine the Berry curvature and the Chern number of the different one-triplon bands. We demonstrate the occurance of Chern numbers $\pm 1$ and $\pm 2$ for the case that two components of the magnetic field are finite. Finally, we also calculate the triplon Hall effect arising at finite temperatures.  
\end{abstract}


\maketitle

%
\section{Introduction}
\label{sec:intro}

The frustrated quantum magnet SrCu(BO$_3$)$_2$ plays an important role in quantum magnetism due to its very rich and complex magnetization curve \cite{kage99,oni00,Kageyama00,Kodama02,Takigawa04,Levy08,Sebastian08,Jaime12,Takigawa13,matsu13}. Experiments in ultrastrong magnetic fields unveil a multitude of intriguing behavior like a series of magnetization plateaus which has triggered a huge body of research over the last years \cite{kage99,oni00,Kageyama00,Kodama02,Takigawa04,Levy08,Sebastian08,Jaime12,Takigawa13,matsu13,miy00,Momoi00a,momoi00,Fukumoto00,Fukumoto01,Miyahara03a,Miyahara03b,dorier08,Abendschein08,Nemec12,Lou12,Corboz2014,Foltin2014,Schneider2016}. Interestingly, the low part of the magnetization curve came into focus only recently, suggesting that SrCu(BO$_3$)$_2$ in a weak magnetic field displays non-trivial topological properties \cite{rom15}, which has been also investigated experimentally by inelastic neutron scattering \cite{clarty16}.

The physical properties of SrCu(BO$_3$)$_2$ can be well described by the Shastry-Sutherland model \cite{shasu81} plus small Dzyaloshinskii-Moriya (DM) interactions \cite{cepas01,rom11}. The non-trivial topological properties then arise from DM interactions being the magnetic analogue of spin-orbit interactions in strongly correlated Mott insulators. As a consequence, the energy bands of the elementary triplon \cite{Schmidt03} excitation of SrCu(BO$_3$)$_2$ in a weak magnetic field can have a finite topological Chern number. The system is therefore expected to be a magnetic version of a Chern insulator \cite{haldane} displaying a triplon Hall effect at finite temperatures \cite{rom15}, similary to other bosonic systems \cite{raghu08,katsura10,zhang10,peano15}.  

In Ref.~\onlinecite{rom15}, Romhanyi and collaborators applied bond-operator theory to an extended Shastry-Sutherland model in the presence of an uniform magnetic field in $z$-direction to investigate the non-trivial topological properties of the triplon bands for realistic values of exchange couplings for SrCu(BO$_3$)$_2$. In this description one has three triplon bands and the triplons behave in momentum space as pseudo-spin one objects coupled to an effective momentum-dependent magnetic field. As a consequence, in a weak magnetic field, topological bands with Chern number $\pm 2$ are found \cite{rom15}.  

In this work we investigate a general uniform magnetic field which is allowed to point in any direction. This is not a trivial extension, since the SU(2) symmetry of the Shastry-Sutherland model is broken due to the DM interactions. Technically, we apply perturbative continuous unitary transformations (pCUTs) \cite{knet00,knet03} about the isolated dimer limit to derive high-order series expansions for the one-triplon bands in the various parameters. Although our calculation does not provide quantitative predictions for the coupling regime relevant to SrCu(BO$_3$)$_2$ in a weak magnetic field, we deduce generic features of the studied system. We show that the more general magnetic field leads to six distinct one-triplon bands and therefore the effective description in terms of pseudo-spin one objects does not hold anymore. As a consequence, there is a richer structure of topological phase transitions as a function of magnetic field with non-trivial Chern numbers $\pm 1$ and $\pm 2$. 

Our paper is organized as follows. In Sect.~\ref{sec:model} we introduce the microcospic model and notations while Sect.~\ref{sec:techni} includes all technical aspects with respect to pCUTs and Chern numbers. Afterwards, we present our results for the topological phase transitions in Sect.~\ref{sec:pd} and we calculate the associated triplon Hall effect in Sect.~\ref{sec:hall}. The main findings are then summarized and discussed in Sect.~\ref{sec:conclusion}. 

%
\section{Model}
\label{sec:model}

\begin{figure}
	\centering
		\includegraphics[width=1.0\columnwidth]{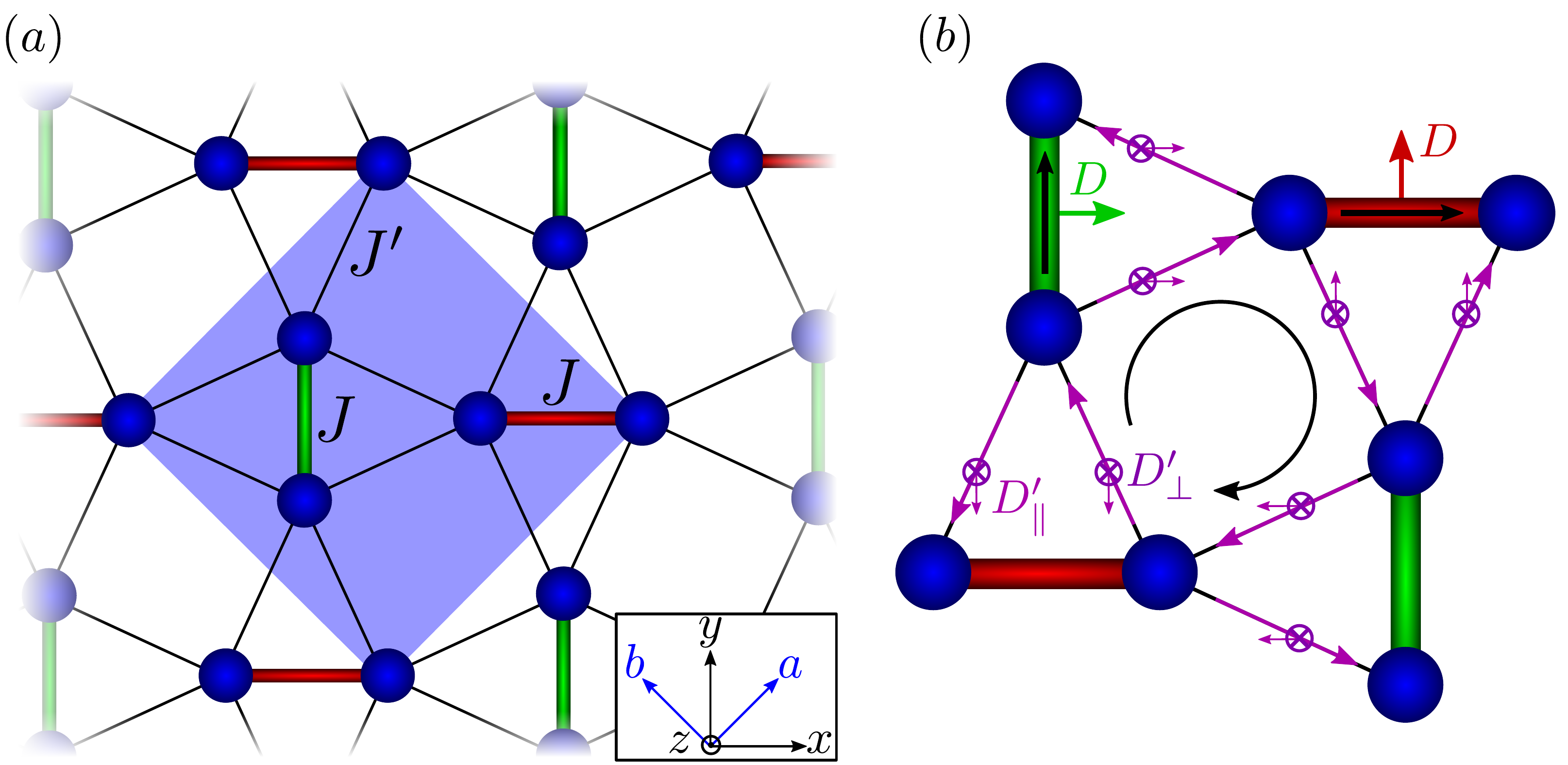}
	\caption{(a) Shastry--Sutherland model consisting of two antiferromagnetic Heisenberg couplings $J$ and $J'$. The Heisenberg couplings within a dimer are denoted by $J$ (horizontal in red and vertical in green) while the inter-dimer couplings are denoted by $J'$. The blue shaded area highlights a primitive unit cell which contains two orthogonal dimers. The corresponding unit vectors $a$, $b$ are depicted in the inset. (b) DM couplings allowed by the symmetries of \ce{SrCu2(BO3)2}. The long arrows on top of the bonds between two spins indicate the orientation of the DM coupling with a corresponding DM vector represented by the short arrow. Accordingly a long arrow from spin $i$ to spin $j$ represent the term $\vec{D} \cdot ( \vec{S}_i \times \vec{S}_j )$. The intra-dimer DM vectors $\vec{D}$ are aligned perpendicular to the dimers. The inter-dimer DM couplings $\vec{D}'$ can be divided into two parts: in-plane components $\vec{D}'_{\parallel}$ and out-of-plane components $\vec{D}'_{\perp}$. The inter-dimer DM vectors are depicted in violet.}
	\label{fig:model}
\end{figure}

We study the same extended Shastry-Sutherland model as in Ref.~\onlinecite{rom15} but in the presence 
of a general uniform magnetic fields $\vec{h}=(h_x,h_y,h_z)$. The specific spin-1/2 Hamiltonian reads
\begin{eqns}
\label{Eq::Ham}
\mathcal{H} &=& J \sum_{\mathrm{n. n.}} \vec{S}_i\cdot \vec{S}_j + J' \sum_{\mathrm{n.n.n.}} \vec{S}_i \cdot\vec{S}_j \nonumber\\
&& {} + \sum_{\mathrm{n. n.}} \vec{D}_{ij}\cdot \left( \vec{S}_i \times \vec{S}_j \right) + \sum_{\mathrm{n. n. n.}} \vec{D}'_{ij}\cdot \left( \vec{S}_i \times \vec{S}_j \right)\nonumber\\
&& - \vec{h}\cdot \sum_{i}\vec{S}_i \quad ,
\end{eqns}
where the first line is the usual Shastry-Sutherland model with antiferromagnetic Heisenberg couplings $J$ and $J'$, the second line contains the intra- and inter-dimer DM interactions, and the last line represents the magnetic field. The various couplings are illustrated in Fig.~\ref{fig:model}. In the following we set $J=1$ throughout the manuscript.

%
\section{Technicalities}
\label{sec:techni}

In this part we give the technical details of our calculation. First, we describe the application of pCUTs about the isolated-dimer limit which allows us to derive the effective one-triplon hopping elements and therefore the calculation of the one-triplon bands. Afterwards, we detail the determination of Chern numbers for these one-triplon bands. 

\subsection{pCUTs}
\label{sec:pcut}
Starting from the isolated-dimer limit, we use a pCUT \cite{knet00,knet03} to derive a quasi-particle conserving effective Hamiltonian up to high order in perturbation along the lines of Refs.~\onlinecite{knet00b,knet04,dorier08,Foltin2014}. In this limit, the unperturbed ground state corresponds to the product state of singlets $\ket{0}\equiv\prod_i\ket{s}_i$ and excitations are local triplets $\ket{t^\alpha}$ with $\alpha\in\{x,y,z\}$. The dimer states are represented by
\begin{eqns}
\ket{s} =& &\frac{1}{\sqrt{2}} \left( \ket{ \uparrow \downarrow } - \ket{\downarrow \uparrow} \right) \\
\ket{t_x} =&\phantom{-}\i &\frac{1}{\sqrt{2}} \left( \ket{ \uparrow \uparrow } - \ket{\downarrow \downarrow} \right) \\
\ket{t_y} =& &\frac{1}{\sqrt{2}} \left( \ket{ \uparrow \uparrow } + \ket{\downarrow \downarrow} \right) \\
\ket{t_z} =& {-\i}&\frac{1}{\sqrt{2}} \left( \ket{ \uparrow \downarrow } + \ket{\downarrow \uparrow} \right) \quad .
\end{eqns}
Next we introduce annihilation and creation operators of triplets $t^{\phantom\dagger}_{i,\alpha}$  and $t^{\dagger}_{i,\alpha}$ on dimer $i$ with flavor $\alpha\in\{x,y,z\}$ so that $t^\dagger_{i,\alpha}\ket{0}\equiv \ket{t^\alpha}_i\prod_{j\neq i}\ket{s}_j$. The unperturbed Hamiltonian of isolated dimers can be written as
\begin{eqnarray}
 \mathcal{H}_0  &=& E_0 + \hat{Q}\nonumber\\
                &=& -\frac{3}{4}N_{\rm d} + \sum_{i,\alpha}\hat{n}_{i,\alpha}\quad ,
\end{eqnarray}
where $N_{\rm d}$ is the number of dimers and $\hat{n}_{i,\alpha}\equiv t^\dagger_{i,\alpha}t^{\phantom\dagger}_{i,\alpha}$. The operator $\hat{Q}$ counts the total number of triplets. The full Hamiltonian \eqref{Eq::Ham} is then expressed as
\begin{equation}
 \mathcal{H} = E_0 + \hat{Q} + \sum_{n=-2}^{n=2}\hat{T}_n \label{Eq:initial_hamiltonian}\quad ,
\end{equation}
where the $\hat{T}_n$ operators increment (or decrement) the number of triplets by $n=\left\{\pm 2,\pm 1,0 \right\}$, i.e.~$[\hat{T}_n,\hat{Q}]=n\hat{T}_n$. 

In total, there are seven perturbations in Eq.~\eqref{Eq:initial_hamiltonian}. Four perturbations act on single dimers. These are the three components $\alpha\in\{x,y,z\}$ of the magnetic field $\mathcal{H}_{\rm \alpha}$ as well as the intra-dimer DM interaction $\mathcal{H}_{\rm D}$. For the magnetic field one has
\begin{equation}
	\mathcal{H}_{\rm x} + \mathcal{H}_{\rm y} + \mathcal{H}_{\rm z} = \hat{T}_0^{\rm x} + \hat{T}_0^{\rm y} + \hat{T}_0^{\rm z}\quad .
\end{equation}
The intra-dimer DM vector $\vec{D}$ is always orthogonal to the (vertical or horizontal) orientation of the dimer as illustrated in Fig.~\ref{fig:model}. The strength of this interaction can therefore be parametrized by a single parameter $D$ and one has
\begin{equation}
	\mathcal{H}_{\rm D} = \hat{T}_1^{\rm D} + \hat{T}_{-1}^{\rm D}\quad .
\end{equation}
The other two perturbations, the Heisenberg interaction $\mathcal{H}_{J'}$ and the inter-dimer DM interaction $\mathcal{H}_{\rm D'}$, couple neighboring dimers. We describe the inter-dimer DM vector $\vec{D}^\prime$ with the two parameters $D'_{\perp}$ and $D'_{\parallel}$ (see also Fig.~\ref{fig:model}). Here $D'_{\perp}$ ($D'_{\parallel}$) refers to the out-of-plane (in-plane) component of the inter-dimer DM interaction. Decomposing the Heisenberg interaction $\mathcal{H}_{J'}$ and the inter-dimer DM interaction $\mathcal{H}_{\rm D'}$ into operator $\hat{T}_n$ yields
\begin{eqnarray}
	\mathcal{H}_{\rm J'} &=& \hat{T}_1^{\rm J'} + \hat{T}_0^{\rm J'} + \hat{T}_{-1}^{\rm J'} \\
        \mathcal{H}_{\rm D'} &=& \hat{T}_2^{\rm D'}+\hat{T}_1^{\rm D'} + \hat{T}_0^{\rm D'} + \hat{T}_{-1}^{\rm D'} + \hat{T}_{-2}^{\rm D'}\quad .
\end{eqnarray}
 
The pCUT then transforms, order by order exactly in the seven perturbations, the initial Hamiltonian \eqref{Eq:initial_hamiltonian} into an effective Hamiltonian $\mathcal{H}_{\rm eff}$ which commutes with $\hat{Q}$: $[\mathcal{H}_0,\hat{Q}]=0$. The effective Hamiltonian therefore conserves the total number of quasi-particles which correspond to triplons\cite{Schmidt03} (dressed triplets) in the current problem. In this work we focus on the one-triplon channel and we calculated all processes up to order seven in all seven perturbative parameters.

\begin{figure*}
	\centering
		\includegraphics[width=1.9\columnwidth]{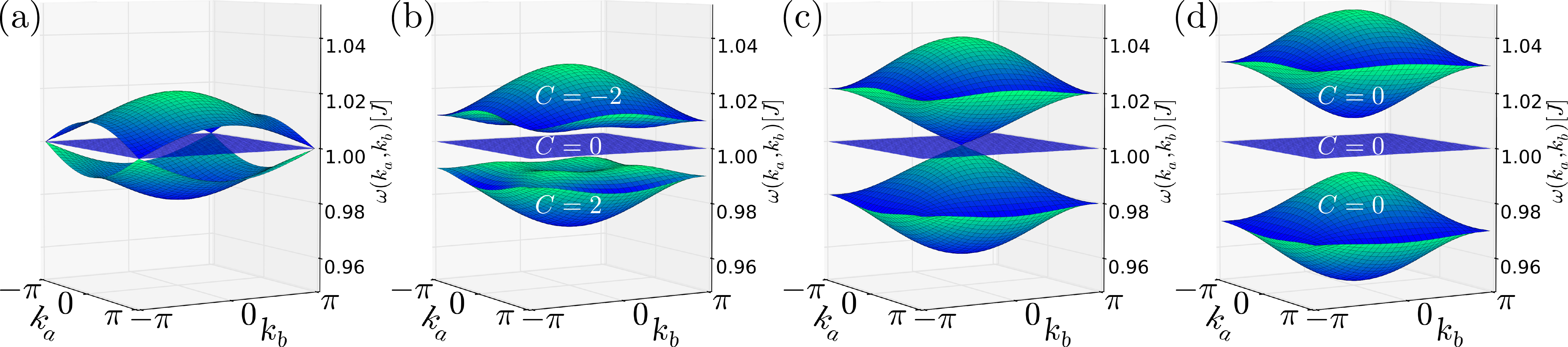}
	\caption{(a)-(d) Evolution of the one-triplon bands $\omega_n ({\bf k})$ in the first-order calculation for $h_x=0$ by tuning the magnetic field from $h_z = 0$ to $h_{z, c}/2$; $h_{z, c}$; $3h_{z, c}/2$ (from left to right) at $D'_{\perp} = 0.01$ and $D'_{\parallel} = 0.007$. The critical point is located at $h_{z, c}=2|D_\perp^\prime|=0.02$ for the chosen parameters. Chern numbers $C$ are given explicitly for (b) and (d).}
	\label{fig:evolution_dispersion}
\end{figure*}

\subsection{One-triplon sector}
\label{ssec:1qp}

All operators of the one-triplon sector correspond to processes where a single triplon 
with flavor $\alpha\in\{x,y,z\}$ hops from dimer $i$ to dimer $i+\delta$ getting the flavor 
$\beta\in\{x,y,z\}$. Consequently, the effective Hamiltonian can be written as
\begin{equation}
	\mathcal{H}_{\rm eff}^{\rm 1qp} = \sum_i\sum_\delta\sum_{\alpha,\beta} a_{\delta}^{\alpha\beta} t^\dagger_{i+\delta,\beta}t^{\phantom\dagger}_{i,\alpha}
\end{equation}
in the one-triplon sector. The hopping amplitudes $a_{\delta}^{\alpha\beta}$ depend on the seven perturbative parameters and they have been calculated by pCUTs as series expansion up to order seven. The Shastry-Sutherland lattice has a two-dimer unit cell as illustrated in Fig.~\ref{fig:model}(a). There are therefore
 six one-triplon states in a single unit cell: the triplon can be either on the vertical (v) or horizontal (h) dimer and in each case it can have a flavor $x$,
$y$, or $z$. Applying Fourier transformation yields then the block-diagonal Hamiltonian
\begin{eqnarray}
\label{H_eff_fou}
 \mathcal{H}_{\rm eff}^{\rm 1qp} &=& \sum_{\bf k} \mathcal{H}_{\rm eff}^{\rm 1qp} ({\bf k})\nonumber\\
                                               &=& \sum_{\bf k} \sum_{\alpha,\beta}\;\sum_{n,m\in\{{\rm v},{\rm h}\}} \omega_{{\bf k}}^{\alpha,\beta,n,m} \, t^\dagger_{{\bf k},\beta,n}t^{\phantom\dagger}_{{\bf k},\alpha,m}\nonumber\\
                                              &=& \sum_{\bf k} {\bf t}^\dagger_{\bf k} \, \tilde{H}^{\phantom{\dagger}}_{\bf k} \, {\bf t}^{\phantom{\dagger}}_{\bf k} \quad ,
\end{eqnarray}
where ${\bf t}^{\phantom{\dagger}}_{\bf k}\equiv (t_{{\bf k},x,v},t_{{\bf k},y,v},t_{{\bf k},z,v},t_{{\bf k},x,h},t_{{\bf k},y,h},t_{{\bf k},z,h})$ and $\tilde{H}_{\bf k}$ is a 6x6 matrix. Diagonalizing $\tilde{H}_{\bf k}$ for all momenta yields the six one-triplon bands $\omega_n ({\bf k})$ so that
\begin{equation}
 \mathcal{H}_{\rm eff}^{\rm 1qp} = \sum_{\bf k} \sum_{n=1}^{6} \omega_n ({\bf k})\, \bar{t}^{\;\dagger}_{{\bf k},n}\bar{t}^{\phantom\dagger}_{{\bf k},n} \quad .
\end{equation}
The explicit matrix elements of $\tilde{H}$ up to order 2 in all perturbation parameters are given in the Appendix \ref{appendix}. Expressions for higher orders become lengthy and can be provided electronically.

\subsection{Chern numbers}
\label{ssec:chern}

The six one-triplon bands $\omega_n ({\bf k})$ are expected to have non-trivial topological properties
 due to the DM interactions, i.e.~they can be characterized by a finite Chern number. The Chern number
 has to be calculated via the one-triplon eigen states $\ket{u_{n{\bf k}}}$ so that $\tilde{H}\ket{u_{n{\bf k}}}=\omega_n ({\bf k})\ket{u_{n{\bf k}}}$. 

If band $n$ is isolated energetically from the other bands, then the Chern number $C_n\in\mathbb{Z}$ is given for this band by
\begin{equation}
 C_n = \frac{1}{2\pi} \iint_{\rm BZ}  F_{n,xy} ({\bf k})\; {\rm d}k_x \; {\rm d}k_y\quad ,
\end{equation}    
where the integral is performed over the Brillioun zone (BZ) and $F_{n,xy}$ is the Berry
 curvature defined by
\begin{equation}
 F_{n,xy} ({\bf k}) \equiv \frac{\partial A_{n,y} ({\bf k})}{\partial k_x} - \frac{\partial A_{n,x} ({\bf k})}{\partial k_y}
\end{equation}   
with
\begin{equation}
 \vec{A}_n({\bf k})={\rm i}\bra{u_{n{\bf k}}}\nabla_{{\bf k}}\ket{u_{n{\bf k}}} 
\end{equation}   
the Berry vector potential.

In a generalized setting a package $p$ of $N>1$ bands is energetically separated from the rest 
of the bands, but the bands included in the package overlap. In this situation one is interested
 in the Chern number $C_p$ of $p$ as an entity which can be calculated as follows. Let $n$ and $m$ be two
 bands being part \smash{of $p$.} The Berry vector potential is then generalized to\cite{Wilczek1984,Mead1992,Soluyanov2012}
\begin{equation}
 \vec{A}_{mn}({\bf k}) \equiv {\rm i}\bra{u_{m{\bf k}}}\nabla_{{\bf k}}\ket{u_{n{\bf k}}} 
\end{equation}     
which allows to define the non-Abelian Berry curvature 
\begin{equation}
 F_{mn,xy}\equiv\mathcal{F}_{mn,xy} - {\rm i}\left[ A_{mn,x},A_{mn,y}\right]
\end{equation}
with
\begin{equation}
 \mathcal{F}_{mn,xy} \equiv {\rm i}\left[\langle \partial_{k_x} u_{n{\bf k}}|  \partial_{k_y} u_{m{\bf k}}\rangle - \langle \partial_{k_y} u_{n{\bf k}}|  \partial_{k_x} u_{m{\bf k}}\rangle\right]\quad . \nonumber
\end{equation}
The Chern number $C_p$ is then obtained by taking the trace over all bands of $p$ and integration over the BZ
\begin{equation}
 C_p \equiv \frac{1}{2\pi} \iint_{\rm BZ}  {\rm Tr}\left[ F_{xy} \right] \; {\rm d}k_x \; {\rm d}k_y\quad .
\end{equation} 
In the following we call a package of bands a {\it multiband} and we refer to $C_p$ as the Chern number of a multiband. Note that we often skip the subscript $p$ ($n$) in $C_p$ ($C_n$) below if there is no ambiguity in order to lighten the notation.

\begin{figure*}
	\centering
		\includegraphics[width=1.9\columnwidth]{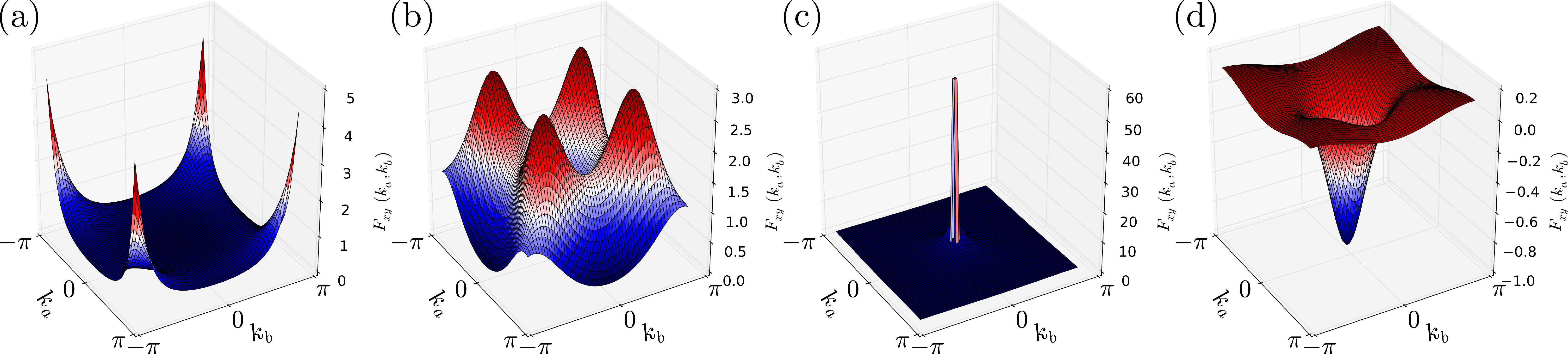}
	\caption{Evolution of the Berry curvature $F_{xy} ({\bf k})$ in the first-order calculation for the lowest multiband at (a) $h_z = 10^{-6}$, (b) $h_{z, c}/2$, (c) $h_{z, c}$, and (d) $3h_{z, c}/2$.}
	\label{fig:evolution_berry}
\end{figure*}

%
\section{Phase diagram in the $h_x$-$h_z$-plane}
\label{sec:pd}
In this part we present our results for the topological phase transitions present in the
extended Shastry-Sutherland \smash{model \eqref{Eq::Ham}} as a function of the magnetic field. We 
 concentrate on the representative parameter set $J' = 0.3$, $D = 0.03$, $D'_{\perp} = 0.01$ 
and $D'_{\parallel} = 0.007 J$ where (i) our bare series from pCUTs are well converged and (ii)
 the relative strengths of the various couplings are appropriate for the frustrated quantum 
 magnet SrCu$_2$(BO$_3$)$_2$ up to a factor two \cite{rom15}. Note that a quantitative 
calculation for SrCu$_2$(BO$_3$)$_2$ demands to go beyond the bare series and needs therefore 
extrapolation schemes for the full six-dimensional band structure including the eigenstates
which is a very challenging task.
Our results are nevertheless generic and expected to be of direct qualitative relevance 
for SrCu$_2$(BO$_3$)$_2$. We find that a magnetic field in the $xy$-plane with $h_z=0$ does
 not introduce gaps in the triplon band structure and therefore does not give any non-trivial
  topological properties. Furthermore, we observe that the sequence of Chern numbers as a function of magnetic field does not depend on the direction of the $xy$-component. As a consequence, we set  $h_y=0$ and study the physical properties in the $xz$-plane.
   
\subsection{$h_z$-field}
\label{ssec:hz}

We start our discussion with the single-field case $h_x=0$. This case has been discussed in 
Ref.~\onlinecite{rom15} for realistic parameters with respect to SrCu$_2$(BO$_3$)$_2$ using a bond-operator treatment inlcuding the first-order effects in the various couplings. Here we would like to show that all generic features concerning the topological nature of the triplon bands are already present for small values $J' = 0.3$. One reason is that the coupling $J'$ does not influence the hopping amplitudes of triplons in first order due to the geometric frustration and the resulting triplon band structure becomes independent of $J'$ as in the treatment from Ref.~\onlinecite{rom15}. 

The evolution of triplon bands as a function of $h_z$ are displayed in Fig.~\ref{fig:evolution_dispersion} using the effective Hamiltonian in first-order perturbation theory (see also Appendix~\ref{appendix}). We find two topological phase transitions
taking place at $h_z = 0$ and \smash{$h_{z, c}=2|D_\perp^\prime|$ \cite{rom15}.} At the quantum critical points, the band structure exhibits a 
three-band touching point located at $M = (\pi, \pi)$ for $h_z = 0$ and at $\Gamma = (0, 0)$ for $h_z = h_{z, c}$. 
This bears a resemblance to Dirac cones with an additional flat band. For $h_z > h_{z, c}$, all triplon bands have
 trivial Chern numbers zero. In contrast, between the two critical points $0 < h_z < h_{z, c}$, the triplon bands 
reveal topological non-trivial Chern numbers $C=\pm 2$ (see Fig.~\ref{fig:evolution_dispersion}(b)). Here the lower 
multiband has $C=+2$ and the upper one $C=-2$. The topological phase transitions and the associated changes of Chern 
numbers can be also seen as divergences in the Berry curvature shown in Fig.~\ref{fig:evolution_berry} 
for the same microscopic parameters. 

\begin{figure}
	\centering
		\includegraphics[width=0.9\columnwidth]{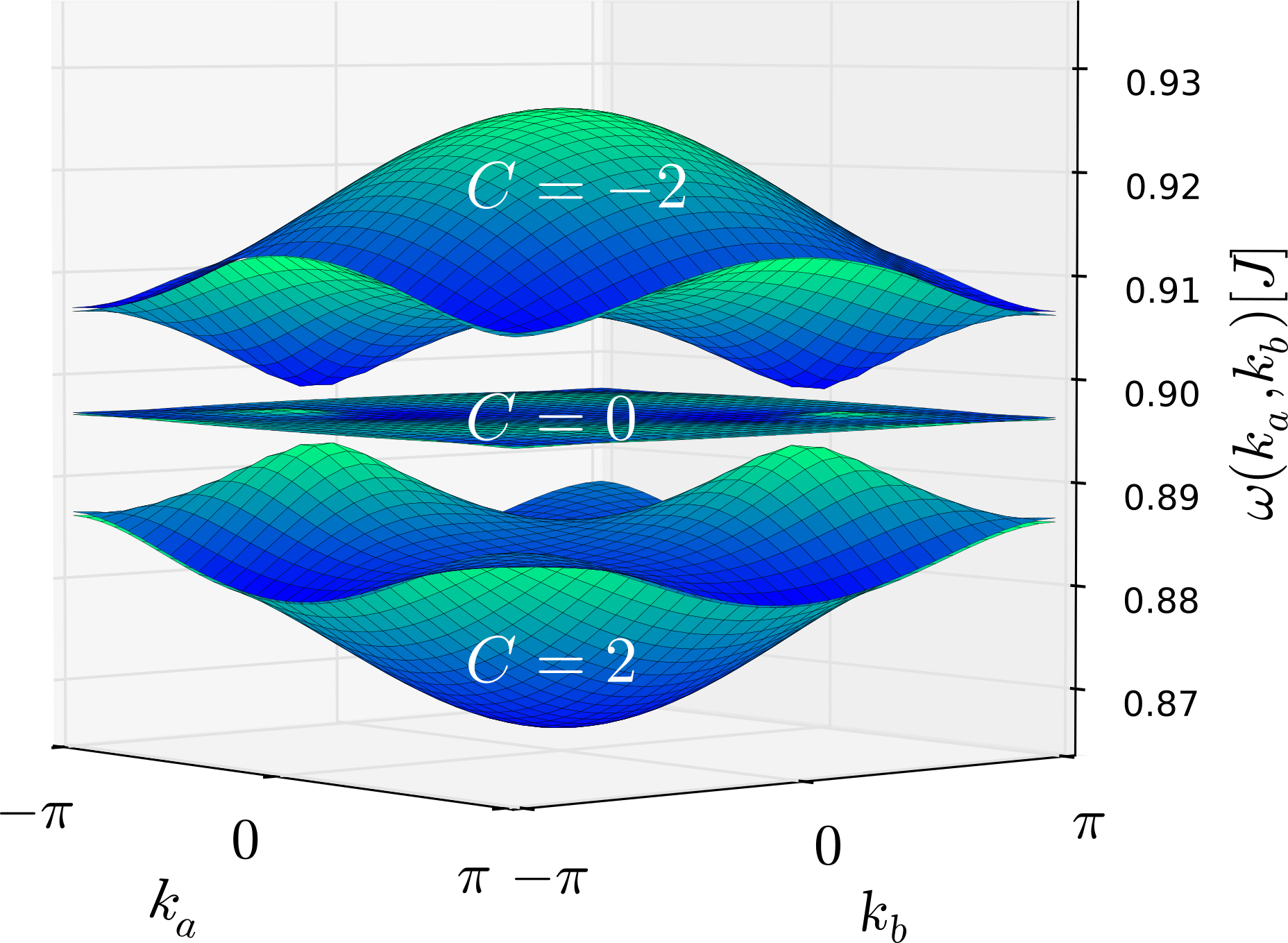} 
	\caption{Dispersion of the one-triplon bands $\omega_n ({\bf k})$ for a magnetic field in $z$-direction with $h_z = 0.01$ at $J' = 0.3$, $D = 0.03$, $D'_{\perp} = 0.01$ and \mbox{$D'_{\parallel} = 0.007$} using the effective Hamiltonian in seventh-order perturbation theory. For these parameters the three triplon multibands have non-trivial Chern numbers: the lowest multiband $C=2$, the middle multiband $C=0$, and the upper multiband $C=-2$.}
	\label{fig:dispersion_hz}
\end{figure}

Inclusion of higher orders in the effective one-triplon Hamiltonian does only result in minor changes in the band structure. As one example, we display the one-triplon band structure inside the topological phase using the effective Hamiltonian in seventh-order perturbation theory in Fig.~\ref{fig:dispersion_hz} for the same microscopic parameters. The three multibands remain unseparated and the middle band is exactly flat in contrast to a magnetic field in $x$-direction discussed below. The overall dispersion has been shifted to lower energies which is the leading (second-order) effect of $J'$ giving a reduction of the local hopping amplitude of triplons.   

\subsection{$h_x$-$h_z$-plane}
\label{ssec:hzhx}
\begin{figure}
	\centering
		\includegraphics[width=0.9\columnwidth]{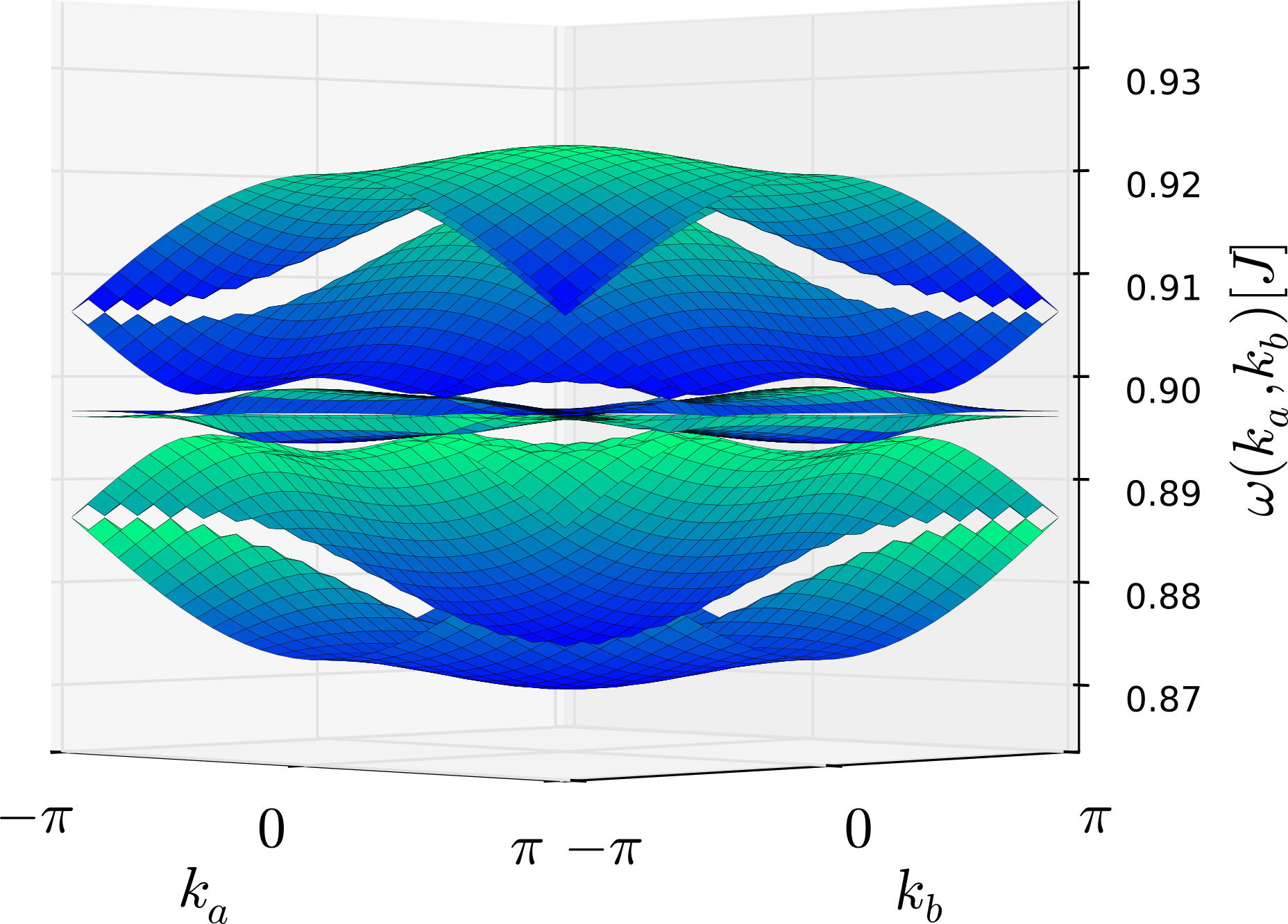} 
	\caption{Dispersion of the one-triplon bands $\omega_n ({\bf k})$ for a magnetic field in $x$-direction with $h_x = 0.01$ at $J' = 0.3$, $D = 0.03$, \smash{$D'_{\perp} = 0.01$} and $D'_{\parallel} = 0.007$ using the effective Hamiltonian in seventh-order perturbation theory. All six bands are connected and therefore no non-trivial Chern numbers are observed.}
	\label{fig:dispersion_hx}
\end{figure}

In this part we study the one-triplon band structure in the $xz$-plane. As mentioned above, a pure field in $x$-direction does not open any gaps between the one-triplon bands and therefore does not induce non-trivial Chern numbers. However, one important difference to the single $z$-field case is the appearance of six one-triplon bands, i.e.~also the two middle bands gain a finite dispersion. As an example, we display in Fig.~\ref{fig:dispersion_hx} the one-triplon bands $\omega_n ({\bf k})$ for a magnetic field in $x$-direction using the effective Hamiltonian in seventh-order perturbation theory.

\begin{figure}
	\centering
		\includegraphics[width=0.9\columnwidth]{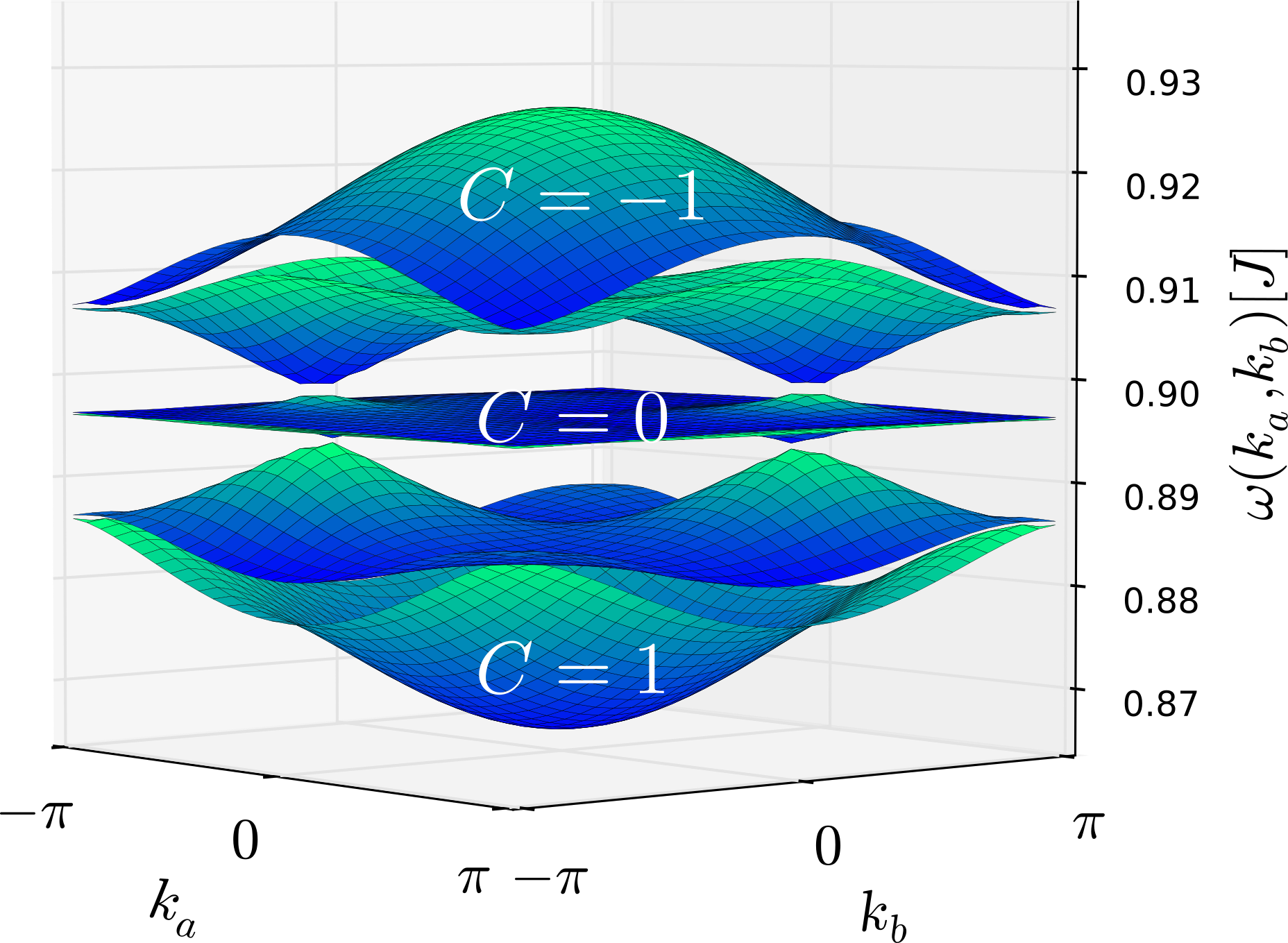} 
	\caption{Dispersion of the one-triplon bands $\omega_n ({\bf k})$ at $h_z = 0.01$ , $h_x = 0.0025$, $J' = 0.3$, $D = 0.03$, $D'_{\perp} = 0.01$ and $D'_{\parallel} = 0.007$ using the effective Hamiltonian in seventh-order perturbation theory. One finds topological non-trivial Chern numbers $C=1$, $C=0$, and $C=-1$ from the lowest to the upper multiband as indicated.}
	\label{fig:dispersion_hz_hx}
\end{figure}

The additional effect of finite $h_x$ and $h_z$ can then result in a richer topological structure of the one-triplon bands. Let us set $h_z=0.01$ and $h_x=0$ so that we are located inside the topological phase where the lowest multiband has Chern number $C=2$ (see Fig.~\ref{fig:dispersion_hz}). If one increases $h_z$ the gap to the middle multiband closes simultaneously at two $k$-points and the Chern number of the lowest multiband jumps to zero. Interestingly, the topological phase transition is different if an additional small field component in $x$-direction is present. In this situation the gaps at the two $k$-points close for slightly different values of the $z$-field and one has two topological phase transitions. The first topological phase transition changes the Chern number of the lowest multiband from $C=2$ to $C=1$ and the second transition from $C=1$ to $C=0$. An example of a triplon band structure with odd Chern numbers is shown in Fig.~\ref{fig:dispersion_hz_hx}.

\begin{figure}
	\centering
	\includegraphics[width=\columnwidth]{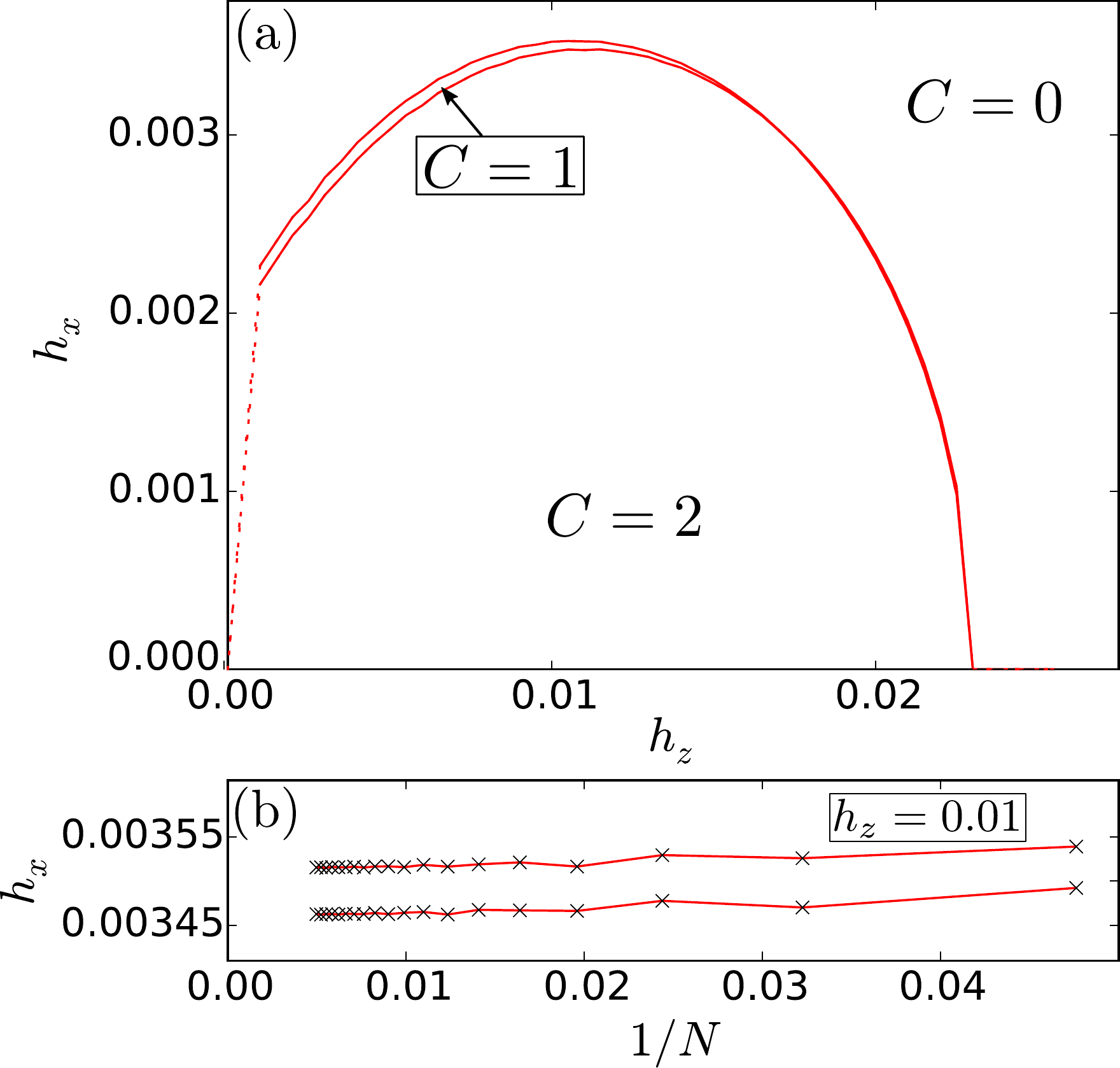}
	\caption{(a) Phase diagram as a function of $h_z$ and $h_x$ at $J' = 0.3$, $D = 0.03$, $D'_{\perp} = 0.01$ and $D'_{\parallel} = 0.007$ using the effective Hamiltonian in second-order perturbation theory. The different phases are characterized by the Chern number of the lowest multiband. For very small values of $h_z$ we were not able to distinguish the two transition lines and we therefore plotted a dashed curve as guide to the eye. (b) The critical fields $h_{x,c}$ versus $1/N$ for a fixed $z$-field component $h_z = 0.01$. Here $N$ refers to the number of wave vectors ${\bf k}$ of the discretized BZ used in the numerical evaluations.}
\label{fig:chernzahlen_hz_hx1}
\end{figure}

The phase diagram in the $xz$-plane characterized by the Chern number of the lowest multiband is shown in Fig.~\ref{fig:chernzahlen_hz_hx1}a for the effective Hamiltonian in second-order perturbation theory. We find that one always has two topological phase transitions with a sequence of Chern numbers $+2$, $+1$, and $0$ for the lowest multiband except the two single-field cases $h_x=0$ and $h_z=0$. The extension of the intermediate phase with an odd Chern number of the lowest multiband is tiny but numericallly robust, e.g.~we show in Fig.~\ref{fig:chernzahlen_hz_hx1}b the influence of the finite wave-vector discretization in the determination of the Chern number.   

\begin{figure}
	\centering
	\includegraphics[width=\columnwidth]{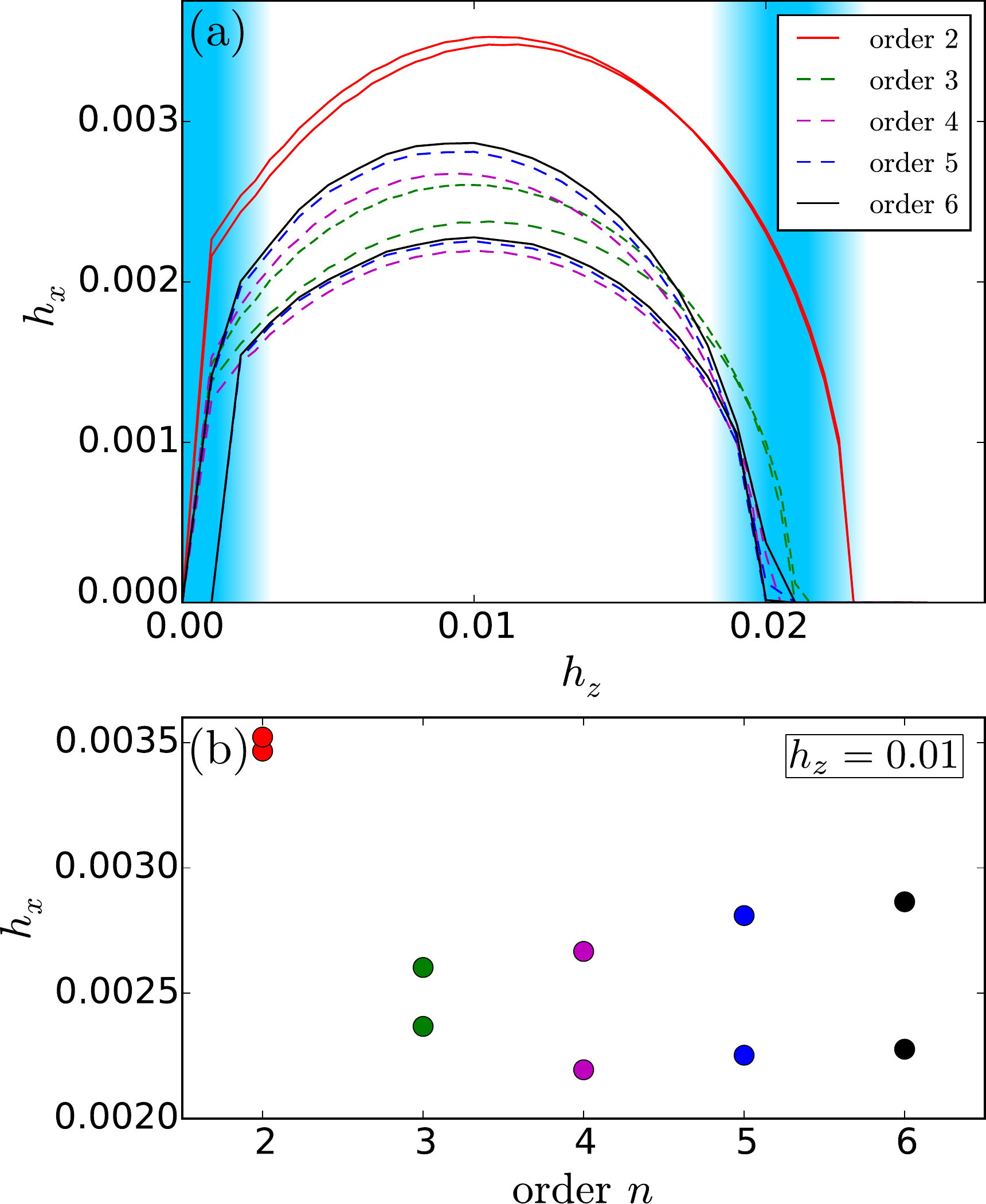}
	\caption{(a) Phase diagram as a function of $h_z$ and $h_x$ at $J' = 0.3$, $D = 0.03$, $D'_{\perp} = 0.01$ and $D'_{\parallel} = 0.007$ for different maximal orders $n$ of the effective Hamiltonian. The different phases are characterized by the Chern number of the lowest multiband. In the shaded regions the differences between the two topological phase transitions can not be resolved numerically. (b) The two critical fields $h_{x,c}$ are depicted versus the perturbative order $n$ for fixed  $z$-field component $h_z = 0.01$.}
\label{fig:chernzahlen_hz_hx2}
\end{figure}

The presence of two distinct topological phase transitions is also robust with the perturbative order as shown in Fig.~\ref{fig:chernzahlen_hz_hx2}. In fact, the intermediate region increases by roughly an order of magnitude compared to the calculation in second-order perturbation theory.

\begin{figure}
	\centering
	\includegraphics[width=0.8\columnwidth]{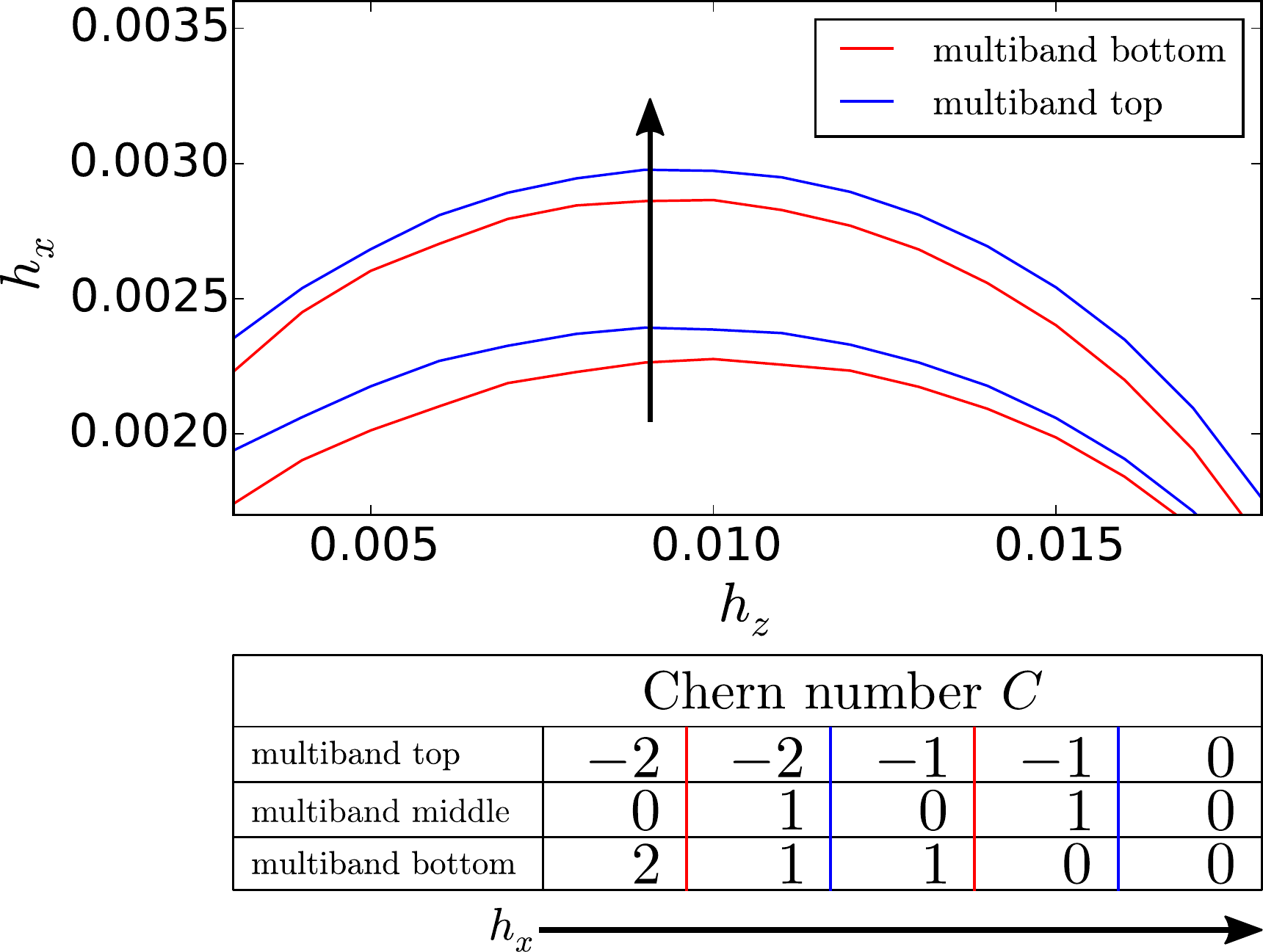}
	\caption{Topological phase transitions with respect to the whole one-triplon band structure using the effective Hamiltonian in sixth-order perturbation theory. The two red curves represent a closing of the gap between the lowest and the middle multiband while the two blue curves signal crossings between the highest and the middle multiband. {\it Lower table}: Evolution of Chern numbers for all three multibands for increasing $h_x$ as marked by the arrow in the figure.}
\label{fig:chernverlauf_hz_hx}
\end{figure}

Finally, we discuss the evolution of the Chern numbers for all three multibands. The corresponding data is displayed in Fig.~\ref{fig:chernverlauf_hz_hx} using the effective Hamiltonian in six-order perturbation theory. We find in total four topological phase transitions as a function of $h_x$ for fixed finite $h_z$. Two phase transitions correspond to gap closings of the lowest multiband as seen above. The other two phase transitions represent level crossings between the middle and the upper multiband. As a consequence, the Chern numbers of the involved multibands change by one as also shown in the table of Fig.~\ref{fig:chernverlauf_hz_hx}.

%
\section{Triplon hall effect}
\label{sec:hall}

 The topological character of Chern insulators in electronic systems are usually detectable by exploiting the integer quantum Hall effect. Due to the non-trivial Chern number topological edge states emerge in an infinite strip geometry with open boundary conditions \cite{hat93}. The edge states are located at the boundary of the sample and have a chiral nature due to the breaking of time reversal symmetry. Furthermore, these gapless edge states connect the conduction and the valence band. If the Fermi level is located in the bulk band gap, which can be done by doping the system, a transverse electrical conductivity results by applying a voltage. The transverse electrical conductivity is then quantized in integer values corresponding to the Chern number \cite{tknn82}.

The Chern number of the one-triplon dispersion gives also rise to gapless chiral edge states \cite{rom15}. But since the triplon excitations are electrically neutral, an applied voltage does not create a different occupation between the oppositely oriented edge states. An alternative approach to detect the topological nature of Chern insulators, which also works for electrical neutral systems, are given by the so-called thermal Hall effect. Applying the Kubo formula in analogy to the electronic case, Matsumoto and Murakami have derived an expression for the thermal Hall conductivity for ordered magnets with magnon excitations \cite{mastumoto11} which also applies for our case of a valence bond solid with triplon excitations. 

Subsequently, Matsumoto and Murakami \cite{mastumoto11_2} showed that the magnon wave packets in insulating magnets have two different rotational motions: a self-rotational motion and a rotational motion along the edge. The two rotational motions of the magnons are caused by the Berry curvature of the magnon bands. The expression for the thermal Hall conductivity in \smash{Ref.~\onlinecite{katsura10}} corresponds to the self-rotational motion. The rotational motion along the edge yields an additional contribution to the thermal Hall conductivity which cancels the self-rotational contribution. The final expression is then completely determined by the rotational motion along the edge. 

To exploit the rotational motion one can create a temperature gradient yielding an anisotropic occupation of the two opposite edge states. The rotational motion of the triplons is no longer balanced and the triplon currents from the two opposite edges do not cancel each other. As a consequence, a finite transverse triplon current appears. This transverse triplon current corresponds to a transverse thermal current, since the triplons are the elementary excitations of the system. Thus, the topological nature of the extended Shastry-Sutherland model can be detected by applying a temperature gradient which leads to a transverse thermal current as already discussed in Ref.~\onlinecite{rom15}. 

\begin{figure}
	\centering
	\includegraphics[width=\columnwidth]{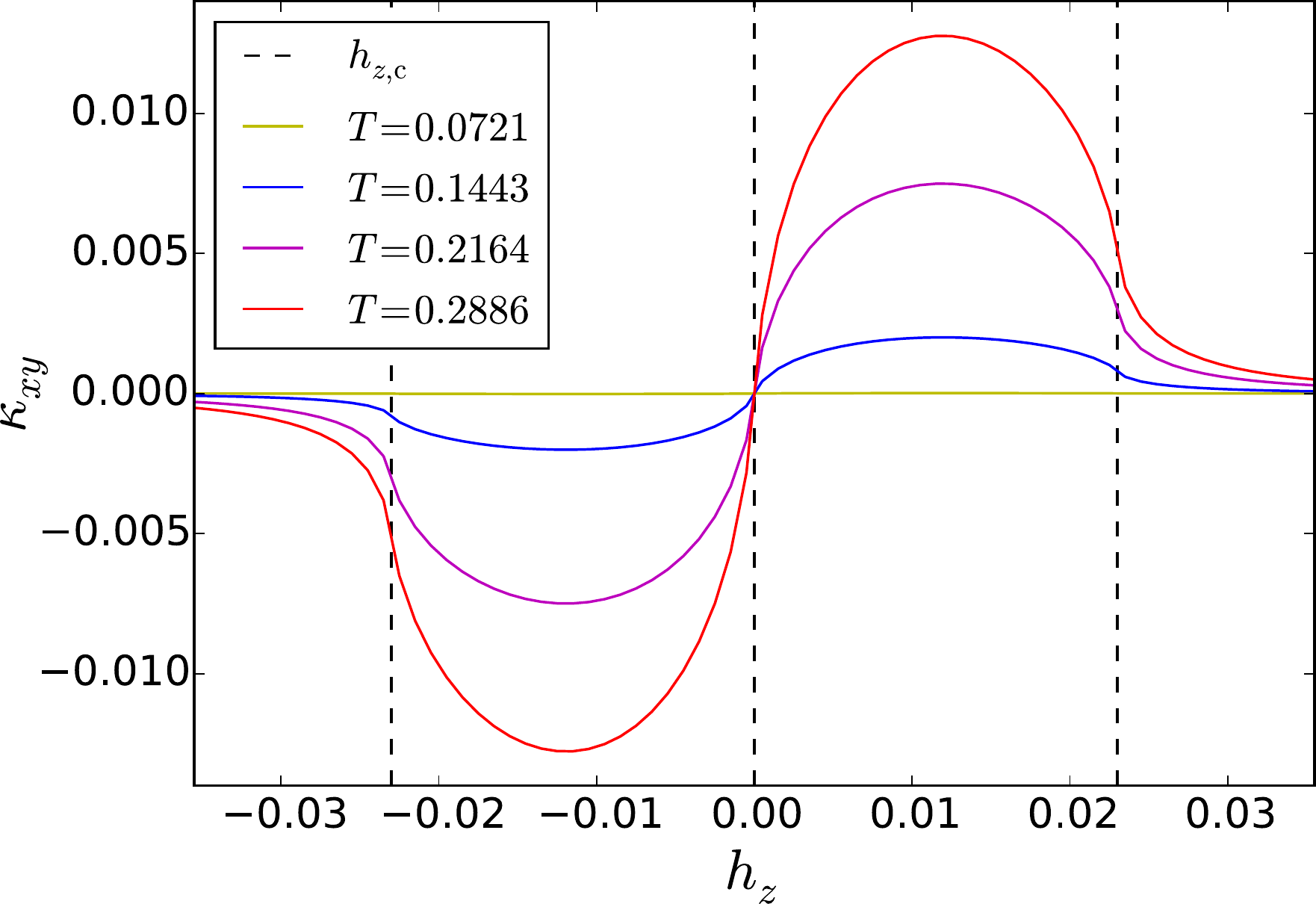} 
	\caption{Thermal Hall conductivity $\kappa^{xy}$ as a function of $h_z$ for various temperatures at $J' = 0.3$, $D = 0.03$, $D'_{\perp} = 0.01$ and $D'_{\parallel} = 0.007$ using the triplon band structure in second-order perturbation theory. The three vertical dashed lines depict the critical value $h_z = \left\lbrace - h_{z, \mathrm{c}}, 0 , h_{z, \mathrm{c}}\right\rbrace$.}
	\label{fig:thermal_hall}
\end{figure}

The expression for the thermal Hall conductivity $\kappa^{xy}$ \cite{mastumoto11_2} is described by 

\begin{eqns}
\kappa^{xy} &=& - T \sum_{n} c_2(\rho_n) F_{n, xy} \\
&=& 2 T \sum_{n} c_2(\rho_n) \operatorname{Im} \Braket{\frac{\partial u_n}{\partial k_{x}} |\frac{\partial u_n}{\partial k_{y}}} \, ,
\end{eqns}
where we measure temperature also in units of $J$ setting \mbox{$k_{\rm B}=1$. \cite{comment_T}} Generally, the quantity $c_q(\rho)$ is defined by

\begin{eqns}
c_2(\rho) &=& \left. \int_{\varepsilon_n}^{\infty} \dx \varepsilon \left( \beta \varepsilon \right)^q \left( - \frac{\dx \rho}{\dx \epsilon} \right) \right|_{\mu = 0} \nonumber \\
 &=& \int_0^{\rho} \left( \log \left( 1 + x^{-1}\right) \right)^{\! q} \dx x \, ,
\end{eqns}

\noindent where $\rho$ is the Bose distribution. For $c_2(\rho)$ we have used the exact expression\cite{mastumoto11_2}

\begin{eqns}
c_2(\rho) &=& - 2 \, \mathrm{Li}_2 (-\rho) + \rho \log^2(\rho^{-1} + 1) - \log^2(\rho + 1) \nonumber \\
 && + 2 \log(\rho + 1) \log(\rho^{-1} + 1) \, , \IEEEeqnarraynumspace
\end{eqns}

\noindent where $\mathrm{Li}_2$ is a polylogarithm function. The topological properties of the extended Shastry-Sutherland model can be tuned by different magnetic fields as shown above. Therefore, we calculate the transverse thermal conductivity $\kappa^{xy}$ in dependence of different magnetic field directions. In the following we focus on the second-order effective Hamiltonian which is also given explicitly in the Appendix \ref{appendix}, since the most important physical properties are already present. First, we investigate the single-field case in $z$-direction. Afterwards, we also calculate the general case of arbitrarily oriented magnetic fields in the $xz$-plane.

\subsection{Pure $h_z$-field}
\label{ssec:hall_hz}

The magnetic field $h_z$ can be used to tune topological phase transitions in the one-triplon bands of the extended Shastry-Sutherland model as seen in the last section. The expression for the Chern number $C$ as well as for the thermal Hall conductivity $\kappa^{xy}$ contains the Berry curvature. Accordingly, one expects that the topological behavior of the system as a function of the magnetic field $h_z$ should be reflected in the thermal Hall conductivity. 

\begin{figure}
	\centering
	\includegraphics[width=\columnwidth]{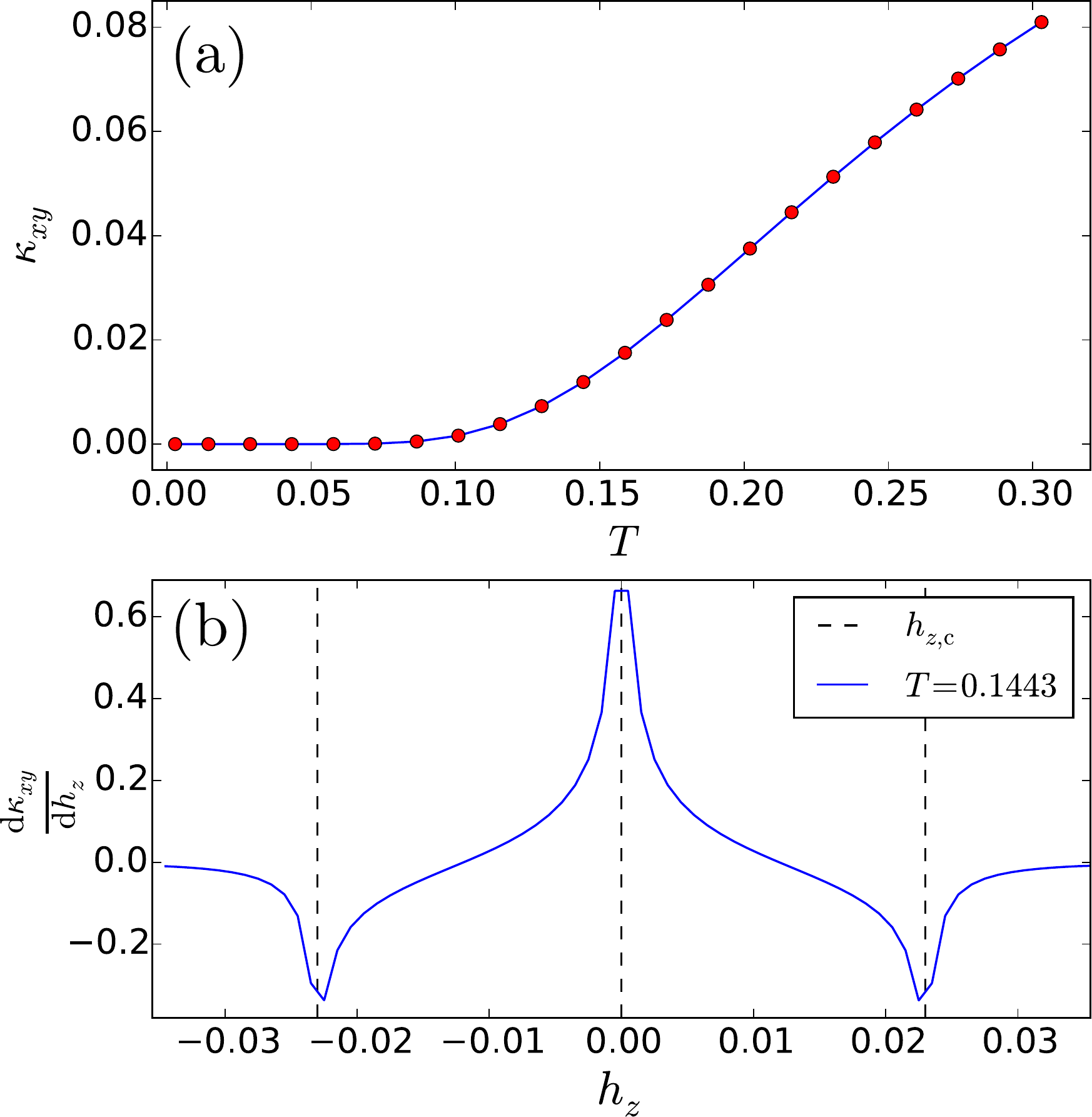}
	\caption{(a) Thermal Hall conductivity $\kappa^{xy}$ as a function of temperature $T$ at $h_z = 0.01$, $J' = 0.3$, $D = 0.03$, $D'_{\perp} = 0.01$ and $D'_{\parallel} = 0.007$. (b) First derivative of the thermal Hall conductivity with respect to $h_z$ as a function of $h_z$ at $T = 0.1443$.}
\label{fig:thermal_hall_prop}
\end{figure}

\begin{figure}[ht]
	\centering
	\includegraphics[width=\columnwidth]{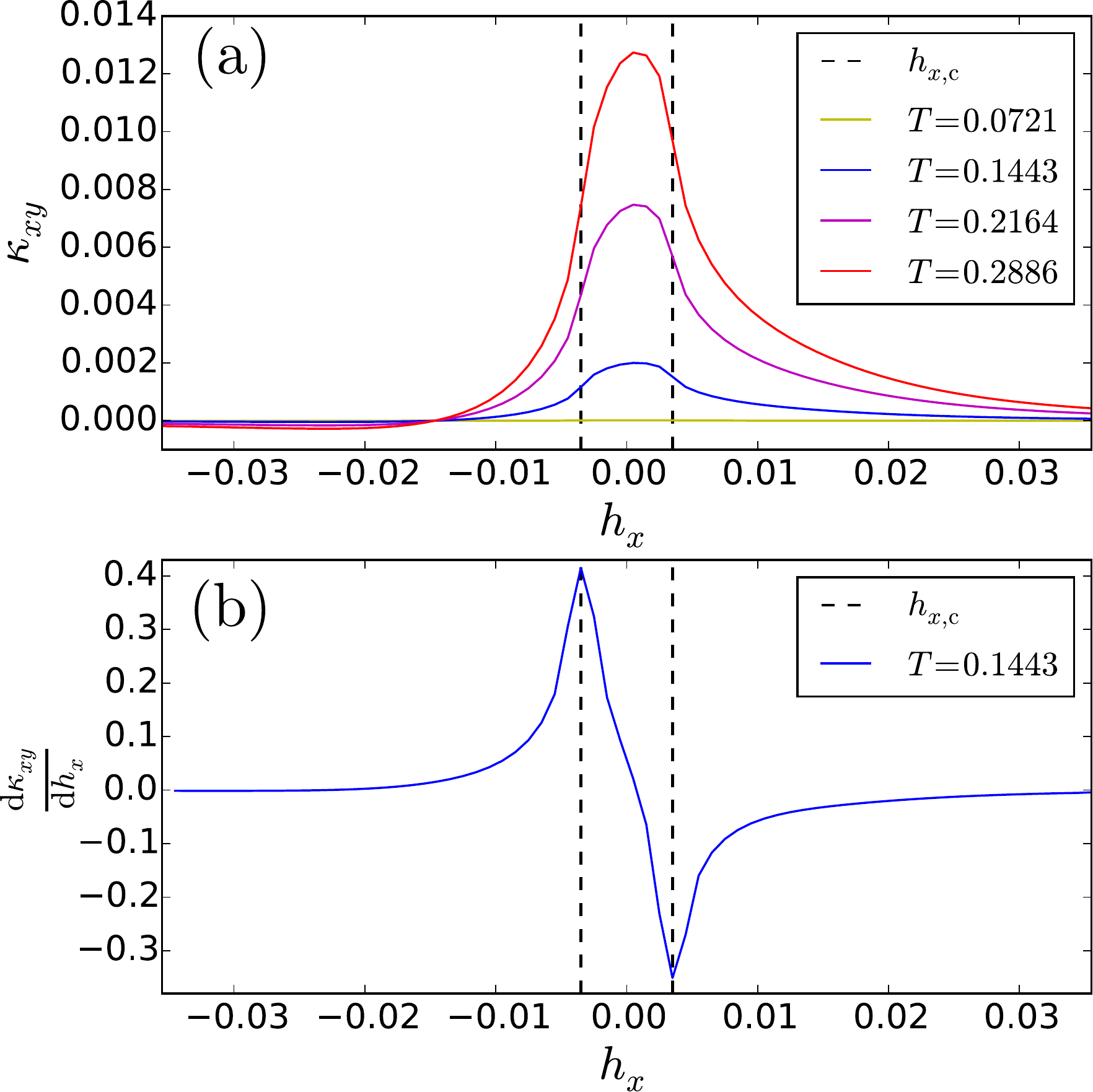}	
	\caption{(a) Thermal Hall conductivity $\kappa^{xy}$ as a function of $h_x$ for various temperatures using the triplon band structure from second-order perturbation theory at $h_z = 0.01$, $J' = 0.3$, $D = 0.03$, $D'_{\perp} = 0.01$ and $D'_{\parallel} = 0.007$. The vertical dashed lines depict the critical value $h_x = \left\lbrace \pm h_{x, \mathrm{c}} \right\rbrace$. (b)  First derivative of the thermal Hall conductivity with respect to $h_x$ as a function of $h_x$ at $T = 0.1443$.}
\label{fig:thermal_hall_hx}
\end{figure}

In Fig.~\ref{fig:thermal_hall} we depict the thermal Hall conductivity as a function of a magnetic field $h_z$ for different temperatures $T$ using the triplon band structure in second order perturbation theory. Calculations to higher order lead only to small modifications. As expected , the largest values for $\kappa^{xy}$ are located inside the topological phase. Turning on the magnetic field leads to non-zero thermal Hall conductivity where the sign is determined by the direction of $h_z$ just like for the sign of the Chern numbers. When the field exceeds the critical magnetic field $h_{z, c}$, the Chern bands become topologically trivial and the thermal Hall signal diminishes. Note that the thermal Hall signal stays finite for $h_z > h_{z, c}$ even if the Chern bands have zero Chern number. This can be traced back to the thermal occupation of the triplon bands leading to different weightings so that equal opposite contributions are no longer canceled exactly and a net non-zero thermal Hall signal is found. We stress that our results for the thermal Hall conductivity show qualitatively the same behavior as the ones calculated for realistic values of $J'$ with respect to the frustrated quantum magnet SrCu$_2$(BO$_3$)$_2$ \cite{rom15}, which is not surprising since the underlying phase diagram is already similar as discussed above. 

Furthermore, it should be noted that increasing the temperature yields higher values for $\kappa^{xy}$. This dependency is displayed in Fig.\ref{fig:thermal_hall_prop}(a). Here we determined the thermal Hall conductivity at $h_z = 0.01$ as a function of temperature. The thermal Hall conductivity is increasing steadily with increasing temperature. The curve can be roughly divided into two linear regimes with two different slopes. The transition point between both regimes is located close to $T = 0.15$. For higher temperatures, the triplon bands are stronger occupied and the triplon-triplon-interaction becomes more and more important, which is beyond the current treatment. 

The topological phase transition points where all three bands form a kind of Dirac cone can not be read off directly from the shape of $\kappa^{xy}$. This is due to the fact that a phase transition does not lead to an instantaneous change of $\kappa^{xy}$. Instead, there are indications that the phase transitions are apparent from the curvature of $\kappa^{xy}$. For this purpose the derivative of $\kappa^{xy}$ with respect to $h_z$ is depicted in Fig.~\ref{fig:thermal_hall_prop}(b). The maxima and minima of the derivative are located at the transition points. This indicates that phase transitions which change the Chern numbers are connected to a change of the curvature.\\

\subsection{$h_x$-$h_z$-plane}
\label{ssec:hall_hzhx}

The topological Chern bands induced by a magnetic \smash{field $h_z$} can also be converted into topological trivial bands by turning on a transverse magnetic field $h_x$. The pure transverse magnetic field $h_x$ can not create a topological phase. Thus $h_x$ has a different topological impact on the system compared \smash{to $h_z$.} Therefore, it is worthwhile to investigate the phase transition induced with $h_x$ by studying the thermal Hall conductivity.

In Fig.~\ref{fig:thermal_hall_hx}(a) we depict the thermal Hall conductivity as a function of a magnetic field $h_x$ at $h_z = 0.01$ for different temperatures using the triplon band structure from second-order perturbation theory. Also in this case the thermal Hall conductivity takes the largest values in the topological phase. The thermal Hall conductivity $\kappa^{xy}$ is not symmetric with respect to the sign of $h_x$ in contrast to the pure $h_z$-case. Switching on $h_x$ leads to a decrease of $\kappa^{xy}$ depending on the direction of the magnetic field $h_x$.

The derivative of $\kappa^{xy}$ with respect to $h_x$ is shown in Fig.~\ref{fig:thermal_hall_hx}(b). The curve contains a maximum and a minimum that indicate the phase transition as in the case of a pure $h_z$ field. These observations are hints that a phase transition is anchored with a change of the curvature of $\kappa^{xy}$. 

\begin{figure}
	\centering
	\includegraphics[width=\columnwidth]{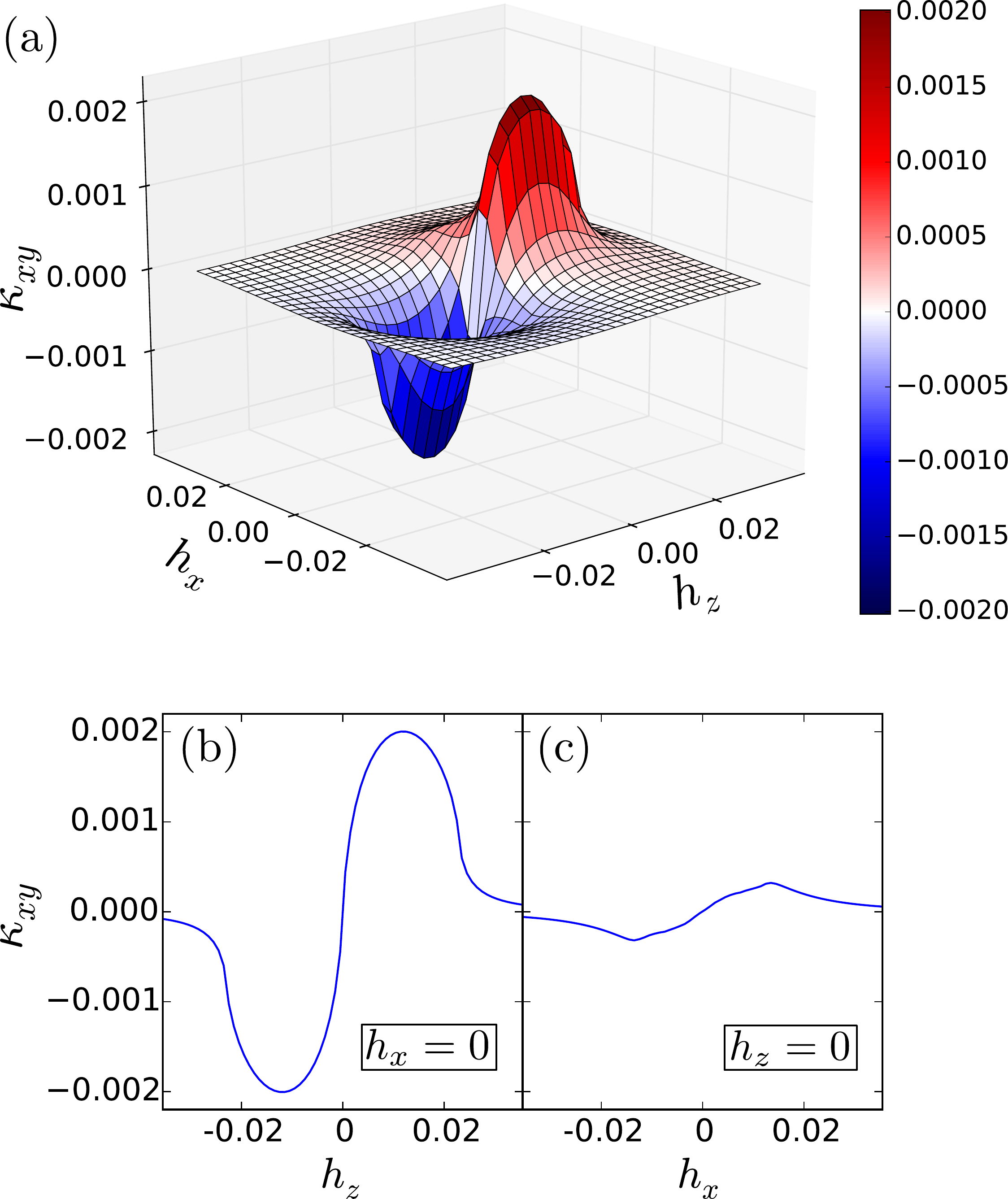}
\caption{(a) Thermal Hall conductivity $\kappa^{xy}$ as a function of $h_z$ and $h_x$ at $T = 0.1443$ using the triplon band structure from second-order perturbation theory at $J' = 0.3$, $D = 0.03$, $D'_{\perp} = 0.01$ and $D'_{\parallel} = 0.007$. (b) and (c) show the limiting cases of a magnetic field pointing in one direction.}
\label{fig:thermal_hall_3d}
\end{figure}

To complete the picture, the thermal Hall conductivity $\kappa^{xy}$ is shown as a function of $h_x$ and $h_z$ in Fig.~\ref{fig:thermal_hall_3d}(a). The largest values for $\kappa^{xy}$ are located on the $h_z$-axis as expected since turning on $h_x$ leads inevitably to the trivial phase. The 3d-plot reveals the point symmetry of $\kappa^{xy}$ at the origin of the coordinate system. The limiting cases of pure magnetic fields in $z$- and $x$-direction are shown in Figs.~\ref{fig:thermal_hall_3d}(b) and (c). The case of a pure $x$-dield shows a non-smooth function due to the fact that the bands are not well isolated. This creates numerical inaccuracies leading to a non-smooth behavior and so the abrupt changes of the curvature do not correspond to phase transitions. Nevertheless, it shows a finite thermal Hall conductivity which is significantly smaller than for the pure $z$-field case.

%
\section{Conclusions}
\label{sec:conclusion}

In this work we have investigated an extended Shastry-Sutherland model which is the relevant
 microscopic model for the frustrated quantum magnet SrCu$_2$(BO$_3$)$_2$. We have used perturbative continuous unitary transformations about the isolated dimer limit to derive
  high-order series expansions for the one-triplon band structure in various couplings 
  including a general uniform magnetic field as well as intra- and inter-dimer DM 
  interactions. 

The main motivation of our study was to extract the non-trivial topological properties of 
the triplon band structure for a general uniform magnetic field extending the investigation of 
 Ref.~\onlinecite{rom15} where a field in $z$-direction has been considered. Although our scheme
 is limited to intermediate ratios $J'/J$ due to the perturbative nature of our series expansion, the qualitative findings are expected to be of relevance for SrCu$_2$(BO$_3$)$_2$. 
 
 We find that a magnetic field in the $xy$-plane does not induce any non-trivial Chern number into the one-triplon band structure. In contrast, if the field has a finite component in $z$-direction and a finite component in the $xy$-plane, then the sequence of Chern numbers and topological phase transitions is richer compared to the pure $z$-case. In the latter the system can be effectively described by a spin-one in a momentum-dependent magnetic field. The band structure consists of three bands which can have Chern numbers $\pm 2$ and $0$. In the more general case where also a finite $xy$-component is present, the Chern numbers of the three multibands can be $\pm 2$, $\pm 1$, and $0$. This is the most important finding of our work.
     
In the future it would be interesting to extend our calculations to larger values of $J'/J$ so that a quantitative modelling of SrCu$_2$(BO$_3$)$_2$ can be achieved. To this end one should go beyond our perturbative treatment and apply non-perturbative variants of continuous unitary transformations \cite{yang11,coester15,krull12,powalski15}. 
       
In addition to the one-triplon band structure, we have also calculated the thermal Hall effect of triplons along the lines of Ref.~\onlinecite{rom15}. The involved approximations are only valid at low temperatures, since triplon-triplon interactions as well as the hardcore constraint have been neglected. A more consistent calculation including these effects are certainly much more challenging, but also very interesting which we leave open for future research. 
 
{\it{Acknowledgement:}}
We thank Judith Romhanyi for fruitful discussions.

%

\appendix
\label{appendix}
\onecolumngrid
\newpage

%
\section{Effective Hamiltonoperator}
\label{sec:effective_ham}

In the following we give the explicit series for the second-order effective Hamiltonian Eq.~\eqref{H_eff_fou}:

\begin{eqns}
\mathcal{H}^{\rm 1qp}_{\mathrm{eff}}({\bf k}) &=&
\begin{pmatrix}
t_{k,x, \mathrm{v}}^{\dagger} \\
t_{k,y, \mathrm{v}}^{\dagger} \\
t_{k,z, \mathrm{v}}^{\dagger} \\
t_{k,x, \mathrm{h}}^{\dagger} \\
t_{k,y, \mathrm{h}}^{\dagger} \\
t_{k,z, \mathrm{h}}^{\dagger} \\
\end{pmatrix}^{\! \top} \! \! \! \!
\begin{pmatrix}
H_{11} & H_{21}^{\dagger} & H_{31}^{\dagger} & H_{41}^{\dagger} & H_{51}^{\dagger} & H_{61}^{\dagger} \\
H_{21} & H_{22} & H_{32}^{\dagger} &  H_{42}^{\dagger} & H_{52}^{\dagger} & H_{62}^{\dagger} \\
H_{31} & H_{32} & H_{33} & H_{43}^{\dagger} & H_{53}^{\dagger} & H_{63}^{\dagger} \\
H_{41} & H_{42} &  H_{43} & H_{44} & H_{54}^{\dagger} & H_{64}^{\dagger} \\
H_{51} & H_{52} &  H_{53} & H_{54} & H_{55} & H_{65}^{\dagger} \\
H_{61} & H_{62} & H_{63} &  H_{64} & H_{65} & H_{66}  \\
\end{pmatrix}
\begin{pmatrix}
t_{k,x, \mathrm{v}}^{\phantom{\dagger}} \\
t_{k,y, \mathrm{v}}^{\phantom{\dagger}} \\
t_{k,z, \mathrm{v}}^{\phantom{\dagger}} \\
t_{k,x, \mathrm{h}}^{\phantom{\dagger}} \\
t_{k,y, \mathrm{h}}^{\phantom{\dagger}} \\
t_{k,z, \mathrm{h}}^{\phantom{\dagger}} \\
\end{pmatrix}
\IEEEeqnarraynumspace \\
H_{11} &=& 1 - J'^2 + \frac{1}{2} D^2 + \left(D'_{\perp} \right)^2 \left(  \frac{1}{2} - \frac{1}{4} \left( \cos(2 k_x)  + \cos(2 k_y)\right) \right. \\
&& \left. - \frac{1}{2} \left( \cos(k_x + k_y) + \cos(k_x - k_y) \right) \right) + \left(D'_{\parallel} \right)^2 \left( \frac{1}{2} + \frac{1}{4} \cos(2 k_y) \right) \nonumber \\
H_{21} &=& - \i h_z + \frac{1}{4} \left(D'_{\parallel} \right)^2 \left( \cos(k_x + k_y) - \cos(k_x - ky) \right)\\
H_{31} &=& \i h_y - \frac{\i}{4} D'_{\perp} D'_{\parallel} \left( \sin(2 k_x) + \sin(k_x + k_y) + \sin(k_x - k_y) \right) \\
H_{41} &=& 0\\
H_{51} &=& - \left( D'_{\perp} + \frac{1}{2} D'_{\perp} J' \right) \left(\cos(k_x) + \cos(k_y)  \right)\e^{- \i k_x}\\
H_{61} &=& - \left( D'_{\parallel} - \frac{1}{2} J' D + \frac{1}{2} J' D_{\parallel} \right) \i \sin(k_y) \e^{- \i k_x} \\
H_{22} &=& 1 - J'^2 + \frac{1}{4} D^2 + \left(D'_{\perp} \right)^2 \left(  \frac{1}{2} - \frac{1}{4} \left( \cos(2 k_x)  + \cos(2 k_y)\right) \right. \\
&& \left. - \frac{1}{2} \left( \cos(k_x + k_y) + \cos(k_x - k_y) \right) \right) + \frac{1}{4} \left(D'_{\parallel} \right)^2  \cos(2 k_x)  \nonumber \\
H_{32} &=& - \i h_x + \frac{\i}{4} D'_{\perp} D'_{\parallel} \left( \sin(2 k_y) + \sin(k_x + k_y) - \sin(k_x - k_y) \right)\\
H_{42} &=& \left( D'_{\perp} + \frac{1}{2} D'_{\perp} J' \right) \left(\cos(k_x) + \cos(k_y)  \right)\e^{- \i k_x} \\
&& + \frac{1}{2} D D'_{\parallel} \left(\cos(k_x) - \cos(k_y)  \right)\e^{- \i k_x} \nonumber \\
H_{52} &=& 0 \\
H_{62} &=& - \left( D'_{\parallel} + \frac{1}{2} J' D + \frac{1}{2} J' D_{\parallel} + D D'_{\perp} \right) \i \sin(k_x)  \e^{- \i k_x}\\
H_{33} &=& 1 - J'^2 + \frac{1}{4} D^2 + \left(D'_{\parallel} \right)^2 \left( \frac{1}{2} + \frac{1}{4} \left( \cos(2 k_x)  + \cos(2 k_y)\right) \right) \\
H_{43} &=& \left( D'_{\parallel} - \frac{1}{2} J' D + \frac{1}{2} J' D_{\parallel} - D D'_{\perp} \right) \i \sin(k_y)  \e^{- \i k_x} \\
H_{53} &=& \left( D'_{\parallel} + \frac{1}{2} J' D + \frac{1}{2} J' D_{\parallel} \right) \i \sin(k_x)  \e^{- \i k_x}\\
H_{63} &=& 0 \\
H_{44} &=& 1 - J'^2 + \frac{1}{4} D^2 + \left(D'_{\perp} \right)^2 \left(  \frac{1}{2} - \frac{1}{4} \left( \cos(2 k_x)  + \cos(2 k_y)\right) \right. \\
&& \left. - \frac{1}{2} \left( \cos(k_x + k_y) + \cos(k_x - k_y) \right) \right) + \frac{1}{4} \left(D'_{\parallel} \right)^2 \cos(2 k_y) \nonumber  
\end{eqns}
\begin{eqns}
H_{54} &=& - \i h_z + \frac{1}{4} \left(D'_{\parallel} \right)^2 \left( \cos(k_x + k_y) - \cos(k_x - ky) \right) \\
H_{64} &=& \i h_y + \frac{\i}{4} D'_{\perp} D'_{\parallel} \left( \sin(2 k_x) + \sin(k_x + k_y) + \sin(k_x - k_y) \right) \\
H_{55} &=& 1 - J'^2 + \frac{1}{2} D^2 + \left(D'_{\perp} \right)^2 \left(  \frac{1}{2} - \frac{1}{4} \left( \cos(2 k_x)  + \cos(2 k_y)\right) \right. \\
&& \left. - \frac{1}{2} \left( \cos(k_x + k_y) + \cos(k_x - k_y) \right) \right) + \left(D'_{\parallel} \right)^2 \left( \frac{1}{2} + \frac{1}{4} \cos(2 k_y) \right) \nonumber\\
H_{65} &=& - \i h_x - \frac{\i}{4} D'_{\perp} D'_{\parallel} \left( \sin(2 k_y) + \sin(k_x + k_y) - \sin(k_x - k_y) \right)\\
H_{66} &=& 1 - J'^2 + \frac{1}{4} D^2 + \left(D'_{\parallel} \right)^2 \left( \frac{1}{2} + \frac{1}{4} \left( \cos(2 k_x)  + \cos(2 k_y)\right) \right)
\end{eqns}


\begin{thebibliography}{99}
\bibitem{comment_T}
Note that the temperature values $T=0.0721$, $T=0.1443$, $T=0.2164$, and $T=0.2886$ translate to $T=2.5\,{\rm K}$, $T=5\,{\rm K}$, $T=7.5\,{\rm K}$, and $T=10\,{\rm K}$ for the absolute value $J=722 \,{\rm GHz}$ used in Ref.~\onlinecite{rom15} for the material SrCu$_2$(BO$_3$)$_2$.
\end{thebibliography}
\end{document}